\documentstyle[12pt]{article}
\newcommand{\be}{\begin{equation}}
\newcommand{\ee}{\end{equation}}
\newcommand{\ba}{\begin{eqnarray}}
\newcommand{\ea}{\end{eqnarray}}
\begin{document}
\renewcommand{\baselinestretch}{1.2}
\small\normalsize
\renewcommand{\theequation}{\arabic{section}.\arabic{equation}}
\renewcommand{\thesection}{\Roman{section}.}
\language0

\begin{flushright}
{\sc BU-HEP} 94-7\\March 1994
\end{flushright}
\vspace{.1cm}

\thispagestyle{empty}

\vspace*{1.5cm}

\begin{center}

{\Large \bf Phase structure and monopoles\\ in U(1) gauge theory}

\vspace*{0.8cm}

{\bf Werner Kerler$^a$, Claudio Rebbi$^b$, and Andreas Weber$^a$}

\vspace*{0.3cm}

{\sl $^a$ Fachbereich Physik, Universit\"at Marburg, D-35032 Marburg,
Germany\\
$^b$ Department of Physics, Boston University, Boston, MA 02215, USA}
\hspace*{3.6mm}

\end{center}

\vspace*{0.5cm}

\begin{abstract}
We investigate the phase structure of pure compact U(1) lattice gauge
theory in 4 dimensions with the Wilson action supplemented by a monopole
term. To overcome the suppression of transitions between the phases in
the simulations we make the monopole coupling a dynamical variable. We
determine the phase diagram and find that the strength of the first order
transition decreases with increasing weight of the monopole term, the
transition thus ultimately getting of second order. After outlining the
appropriate topological characterization of networks of currents lines,
we present an analysis of the occurring monopole currents which shows
that the phases are related to topological properties.
\end{abstract}

\newpage

\section{Introduction}
\setcounter{equation}{0}
\hspace{0.35cm}
The investigation of the phase structure of U(1) lattice gauge theory is
important in two respects. First, the theory should be the basis of
QED which is still not understood at the nonperturbative level. Second,
it provides a unique model to study the interrelation between phase
structure and topological properties of the field configurations.

The phase transition in 4D compact U(1) lattice gauge theory is known
to be related to the occurrence of monopoles. Using the formulation
of DeGrand and Toussaint \cite{dt80} Barber et al.~have shown that,
if one adds a monopole term to the action, depending on its weight
the transition can be suppressed \cite{bss85} or shifted \cite{bs85}.
The consequences of removing monopoles in the U(1) theory
have also been studied in \cite{bmm92}.

Recent results \cite{acg91,blsu92} on the theory without a monopole term
give energy histograms which indicate a first-order transition. The
problem in these simulations is that the tunneling between the phases
is strongly suppressed. In order to overcome the difficulty
the authors of  \cite{acg91} introduce some type of iterative reweighting
for different $\beta$, while the authors of \cite{blsu92} use a matching
of hot and cold start results.

To reconsider the case with a monopole term is of interest in two
respects. First the variation of details of the transition with the
weight of this term provides further insight into the properties of the
theory. Second, if the strength of the transition turns
out to decrease with this weight, then, by making
the weight a dynamical variable, one can set
up a very efficient simulation algorithm. The efficiency of such
procedure  has
been demonstrated in \cite{kw93},
where, by making the number of states $q$ in the Potts
model a dynamical variable, the authors could bridge the
energy gap that occurs for $q>4$.

A further aspect of the U(1) theory
which deserves reconsideration are the properties of
the spatial structure of the monopole currents in the
configurations and their relationship to the global features
showing up in the observables. First results along these lines have been
presented some time ago by Gr\"osch et al.~\cite{gjjlnr85}. Recently
this issue has been addressed again by Bode et al.~\cite{bls93} who
have observed that in 4D compact U(1) theory one is confronted with
clusters of monopole lines rather than with single loops.

Because the topological description is straightforward for loops only,
this rises the question if a satisfactory topological characterization
of networks of current lines can be found. In this context one should also
discuss the work by Lang and Neuhaus \cite{ln93}, who, simulating
the model on the surface of a 5D hypercube (homeomorphic to a 4-sphere)
rather than on
a 4D lattice with periodic boundary conditions (corresponding to a
torus), found that the first order signal disappears.

In the present paper we show that the additional monopole term provides
the features needed to set up a powerful simulation algorithm. Our
investigations give detailed properties of the phase transition and their
dependence on the weight of the monopole term. We
introduce a topological characterization of networks
of current lines and find that the
topology of these networks signals the phases.

Sec.~II gives definitions and general relations. In Sec.III the
method of the Monte Carlo simulations is described. Sec.~IV
presents the results based on histograms and a discussion of the
phase structure. In Sec.~V the topological characterization of
networks of current lines is outlined. In Sec.~VI the numerical
results on current networks are presented and discussed. Sec.~VII
contains some conclusions.

\section{Action and monopole currents}
\setcounter{equation}{0}
\hspace{0.35cm}
The Wilson action supplemented by a monopole term is of form
\be
S=\beta \sum_{\mu>\nu,x} (1-\cos \Theta_{\mu\nu,x})+
\lambda \sum_{\rho,x} |M_{\rho,x}| \quad .
\label{sbl}
\ee
In terms of the link angles $\Theta_{\mu,x}\in [-\pi,\pi)$ the plaquette
flux $\Theta_{\mu\nu,x}\in (-4\pi,4\pi)$ is
$\Theta_{\mu\nu,x}=\Theta_{\mu,x}+\Theta_{\nu,x+\mu}-\Theta_{\mu,x+\nu}-
\Theta_{\nu,x}$. The physical flux $\bar{\Theta}_{\mu\nu,x}\in [-\pi,\pi)$
is defined \cite{dt80} by
\be
\Theta_{\mu\nu,x}=\bar{\Theta}_{\mu\nu,x}+2\pi n_{\mu\nu,x}
\ee
where $n_{\mu\nu,x}=0,\pm 1, \pm 2$. The monopole content of 3D cubes
which enters the additional term in (\ref{sbl}) is given by
\be
2\pi M_{\rho,x} = \frac{1}{2}\epsilon_{\rho\sigma\mu\nu}
(\bar{\Theta}_{\mu\nu,x+\sigma}-\bar{\Theta}_{\mu\nu,x})
\ee
where $M_{\rho,x}=0,\:\pm 1,\: \pm 2$. The $\bar{\Theta}_{\mu\nu,x}$ are
invariant under gauge transformations $\Theta'_{\mu,x} = [\Theta_{\mu,x}
+\chi_{x+\mu}-\chi_x+\pi] \mbox{ mod } 2\pi -\pi$ with $\chi\in [-\pi,\pi)$
(which guarantees that $\Theta'_{\mu,x}\in [-\pi,\pi)$ as well).

We find it convenient to introduce $J_{\rho,x}=M_{\rho,x+\rho}$ and
$\vartheta_{\rho\sigma,x}= \frac{1}{2}\epsilon_{\rho\sigma\mu\nu}
\bar{\Theta}_{\mu\nu,x+\rho+\sigma}$ because then the current
conservation law
\be
\sum_{\rho}(J_{\rho,x}-J_{\rho,x-\rho})=0
\label{cc}
\ee
and the field equation
\be
\sum_{\sigma}(\vartheta_{\rho\sigma,x}-
\vartheta_{\rho\sigma,x-\sigma}) = 2\pi J_{\rho\sigma}
\label{fe}
\ee
have a straightforward geometric interpretation on the dual lattice.

Summing (\ref{fe}) over three of the four coordinates, for
periodic boundary conditions we obtain
\be
\sum_{x_{\mu_0}x_{\mu_1}x_{\mu_2}}J_{\mu_3,x}= 0 \quad ,
\label{nf}
\ee
i.e.~the vanishing of the net current flow through any hypersurface
perpendicular to the direction of the flow. We note that (\ref{cc}) holds
separately on each network ${\bf N}$ of current lines (cf. Sec.~V)
disconnected from the rest. Therefore, summing (\ref{cc}) in that case
over three of the four coordinates we still get
\be
\sum_{x_{\mu_0}x_{\mu_1}x_{\mu_2}}J_{\mu_3,x}= f_{\mu_3} \quad
\mbox{for} \quad J_{\mu,x}\in {\bf N}
\label{nfn}
\ee
where $f_{\mu_3}$ is constant. By (\ref{nf}) the net current
flows $f_{\mu}$ of the
occurring networks  have to sum up to zero.

\section{Method of simulation}
\setcounter{equation}{0}
\hspace{0.35cm}
In the usual simulations $\lambda$ is a fixed parameter and one deals
with a probability distribution $\mu_{\lambda}(\Theta)=
\exp(-S_{\lambda}(\Theta))/Z_{\lambda}$. In order to make the parameter
$\lambda$ a dynamical variable we consider $\mu_{\lambda}(\Theta)$
as the conditioned probability to get a configuration $\Theta$ given a
definite value
$\lambda$ and allow the values of $\lambda$
to vary with a chosen probability distribution $f(\lambda)$. Then, to
simulate the joint probability distribution $\mu(\Theta,\lambda)=
f(\lambda)\mu_{\lambda}(\Theta)$, we use  $\mu(\Theta,\lambda)=
\exp(-S(\Theta,\lambda))/Z$ with $S(\Theta,\lambda)=S_{\lambda}(\Theta)
+g(\lambda)$. This implies that the relation between $f$ and $g$ is
given by
\be
f(\lambda)=Z_{\lambda} \exp(-g(\lambda))/Z
\label{fg}
\ee
with $Z=\sum_{\lambda}Z_{\lambda}\exp(-g(\lambda))$.

In the simulations we use a discrete set of $n$ values of $\lambda$. For
the efficiency of the simulations an appropriate choice of $f(\lambda)$
is crucial. We require this to be (approximately) constant so that
(almost) identical
total numbers of sweeps are spent at all values of $\lambda$. By
(\ref{fg}), constant
$f(\lambda)$  means $g(\lambda)=\ln Z_{\lambda}+c$, with an
arbitrary constant $c$. Reasonable values of $g(\lambda)$ are readily
obtained by short runs at fixed $\lambda$. These values can be improved
iteratively in the full simulations, replacing $g(\lambda)$ by
$g(\lambda)+\ln(nf(\lambda))$ in subsequent iterations, which
converges to constant $g(\lambda)$.

While the total amount of time spent at a definite $\lambda$ value is fixed by
$f(\lambda)$, there is still the freedom to vary the average stay time (the
average number of sweeps spent at a particular $\lambda$ before leaving it).
The reciprocal of this time is the sum of the transition probabilities
to the neighboring $\lambda$ values. To get an efficient algorithm these
probabilities must not be too small, which means that
one must use a sufficient number
of $\lambda$ values. Further it appears appropriate to make
these probabilities (roughly) the same in all cases, which can be achieved
by adjusting the distances between the $\lambda$ accordingly.

In our application of the algorithm each update of the $\Theta$
link variables has
been followed by an update of $\lambda$. As individual update steps
we have used Metropolis steps in both cases. For the $\lambda_q$ with
$q=1,\ldots,n$ we have used the proposal matrix $\frac{1}{2}(\delta_{q+1,q'}
+\delta_{q,q'+1}+\delta_{q,1}\delta_{q',1}+\delta_{q,n}\delta_{q',n})$.

For the efficiency of the algorithm it is crucial that
the fact of making $\lambda$ a dynamical variable
opens an easier pathway between the phases. In the case
under consideration
this happens because the strength of the first-order
transition decreases with $\lambda$. This is illustrated by Figure 3
(discussed in Sec.~IV), which indicates that traveling along the peaks is
easier than tunneling through the valleys.

For each value of $\lambda$ one must fix a corresponding value
of $\beta$.
To exploit the algorithm already at the stage when one searches for the
transition line, the $\lambda$ interval can be gradually extended,
starting from a region with overlapping peaks and adjusting the
$\beta$ values with increasing statistics.

The tunneling times \cite{bp91} between the phases for our algorithm
with dynamical $\lambda$ are greatly reduced as compared to those of a
conventional Metropolis algorithm. For example, for $L=8$ at the
phase transition line (cf. Sec.IV) we get, in units of $10^3$, 0.17(2)
as compared to 3.0(3) for $\lambda =0$ and 0.25(4) as compared to
23(5) for $\lambda =-0.3$.

Since we are interested in the results
for all of the $\lambda$ values considered, these
times reflect the actual gain. If we were interested only in one of
the $\lambda$ values, to make a fair comparison
we would have to
multiply the times of our algorithm roughly by the number of $\lambda$
values used. Thus envisaging interest in the results
for $\lambda=-0.3$ only, multiplying the above timing by $n=21$,
we would find that there remains still
considerable gain. Envisaging only $\lambda=0$, where
traveling to negative values is no longer needed, the factor would be 15
and there would be some gain there as well.

The actual point, however, is that for large $L$ (in which one is mainly
interested), and similarly also at negative $\lambda$, our algorithm is in
any case superior. This occurs simply because then the peaks in the
energy distribution related to the phases (cf.~Sec.~IV) get separated
so that by using conventional algorithms one does not observe any
transitions at all.

A further important virtue of the present algorithm is that it allows
vectorization and parallelization of the computer programs (as does
e.g.~not hold for the multicanonical method \cite{bn91}). This has
allowed us to develop an efficient parallel implementation running on
the connection machine CM-5.

In the present work we have run the algorithm with dynamical $\lambda$
for $L=8$ using the 21 values of $\lambda$, and corresponding values of
$\beta$ and $g(\lambda)$, given in Table I. For $L=8$ at $\lambda=0.9$ and
for $L=16$ at $\lambda=0$ and at $\lambda=0.6$ we have also
performed conventional simulations
at a number of $\beta$ values. The statistics we
collected is larger than $10^5$ sweeps for each of the $\lambda$ values
we considered for $L=8$ as well as for $L=16$.

\section{Phase structure}
\setcounter{equation}{0}
\hspace{0.35cm}
Rather precise results on the phase transition
have recently been obtained \cite{acg91,blsu92} in the absence of a
monopole term by considering energy histograms. Including
the monopole term in investigations based on histograms,
we find that the strength of the first order transition decreases as
$\lambda$ increases. This is seen from Figure 1, which compares the
distributions $P(E)$ of the average plaquette energy
$E=(1/6L^4)\sum_{\mu>\nu,x} (1-\cos \Theta_{\mu\nu,x})$
on lattices of sizes $L=8$ and $L=16$ for $\lambda=0$ and $\lambda=0.6$~.

As the peaks overlap the determination of the location of the phase
transition needs special care. This is illustrated by Figure 2, which for
$\lambda=0.6$ and $L=16$ shows the sensitivity to the value of $\beta$.

 From Figure 3, which presents the distribution $P(E,\lambda)$ we obtained in
the transition region for $L=8$ by simulations with dynamical $\lambda$,
the decrease of the strength of the transition is seen in more detail.
The figure also makes the case for our algorithm: from the profile of the
distribution it is clear how a simulation with dynamical $\lambda$
can trace the peaks and thus avoid the long correlation times
due to the separation of the phases in the region where the transition
is strongly of the first-order.

We define as location of the phase transition the maximum of the specific
heat, which determines for us
$\beta_{\mbox{\scriptsize{C}}}$ for given $\lambda$. To
adjust the data measured in the transition region appropriately, we use
reweighting \cite{fs88}. The location of the transition in $(\beta,\lambda)$
space is depicted in Figure 4 for $L=8$. Corresponding numerical values of
$\beta_{\mbox{\scriptsize{C}}}$ up to $\lambda=0.6$ are given in Table I.
For $L=8$ at $\lambda=0.9$ we get $\beta_{\mbox{\scriptsize{C}}}=0.3885(5)$
and for $L=16$ at $\lambda=0.6$ we obtain $\beta_{\mbox{\scriptsize{C}}}=
0.6428(3)$; the value for $L=16$ at $\lambda=0$ determined in
Ref.~\cite{blsu92} is $\beta_{\mbox{\scriptsize{C}}}=1.01082(6)$.

Figure 5 presents the latent heat as a function of $\lambda$ at the phase
transition line for $L=8$, confirming the fact, already seen from Figure 1
for $L=8$ and for $L=16$ and from Figure 3 for $L=8$ for a whole range
of $\lambda$ values, that the strength of the transition decreases with
increasing $\lambda$. We give the data in Figure 5 up to the point where
separation of phases appears numerically justified. The merging of the
peaks signals that the transition ultimately gets of second order (a further
indication of this will be discussed in Sec.~VI).

We confirm the observation \cite{blsu92} for $\lambda=0$ that the latent
heat at the transition point from $L=8$ to $L=16$ decreases. However,
the requirement for
a first order transition is only that
extrapolation to infinite $L$ of the latent heat versus $1/L$
leads to a finite value \cite{lk90}.  For the Potts model it
has recently been demonstrated that such an extrapolation does reproduce known
results \cite{j93}. If we extrapolate our values 0.046(2) for $L=8$
and 0.030(2)
for $L=16$ linearly versus $1/L$ we obtain the finite value 0.014. It
is, nevertheless, still an open question to what extent finite
size scaling already
applies.

Our results for the distributions of the monopole number density $\rho=
(1/4L^4)\sum_{\rho,x} |M_{\rho,x}|$ are very similar to the ones presented
above for the distributions of the plaquette energy $E$. This confirms the
strong correlation between $E$ and $\rho$ at the transition point which has
been known for some time \cite{b84}. In Figure 6, which illustrates
data obtained with
$L=8$ and $\lambda=-0.3$, we show the preferred direction of
the distribution $P(E,\rho)$ in $(E,\rho)$ space. From Figure 7,
which exhibits $P(E,\rho)$ for
$L=8$ and $\lambda=-0.3,\:0,\:0.3,\:0.6,\:0.9$, it is seen that the slope of
the correlation shows only little dependence on $\lambda$. In addition from
Figure 7 it is apparent that the $\rho$ of the cold phase
is roughly the same for
all $\lambda$.

 From Figure 7 we see that the slope of the correlation between
plaquette energy and monopole density $\Delta E / \Delta \rho$
ranges from approximately $1.1$ to approximately $1.3$, with the smaller
value slightly favored for larger monopole density.  Remembering
that the ratio between total numbers of plaquettes and cubes in a
four-dimensional lattice
is $3/2$, this indicates that the average total extra plaquette energy
associated with the presence of a monopole is
$\Delta E_{tot} \approx 1.8 $.  A semi-classical explanation for this
number can be obtained along the following lines.  A calculation of the
minimal plaquette energy needed to produce a monopole loop of length
$4$ in an otherwise totally ordered field configuration gives
$ E = 6.65 $ i.e. a total plaquette energy per monopole
$\approx 1.61 $. (We have used a constrained relaxation technique
to evaluate this number. One must of course impose a constraint
since a monopole loop is classically unstable.)
This can account for the value of $\Delta E / \Delta \rho$
in the low monopole density regime.  With a very high density
of monopoles a  more appropriate quantity to consider would be
the total plaquette energy necessary to produce a long monopole line.
In this case the plaquette energy per monopole can be obtained by
calculating the total plaquette energy for
a single monopole configuration in a three-dimensional system,
which is given by $ E = 4.41 $.  This
number is much larger than the observed $\Delta E / \Delta \rho$,
but one must also consider that in  regimes of high monopole densities
the monopoles are produced over the backgound of a rather disordered gauge
field. Thus it would not seem correct to attribute to the presence
of a monopole the entire energy necessary to create it from the vacuum,
but instead only the excess energy over the background.  For this
reason we have also calculated the excess plaquette energy above some definite
cutoff in a classical monopole configuration (i.e., we have summed
$\min (E_{plaquette}-E_{cutoff}, 0)$ over all plaquettes).  With
$ E_{cutoff}=0.2 $ and $0.25$ this gives $E = 1.80$ and $1.50$
respectively.  Thus, in either case (low and high density of monopoles) the
number that emerges from the semiclassical calculation is in reasonable
agreement with the observed values.

\section{Characterization of networks}
\setcounter{equation}{0}
\hspace{0.35cm}
The currents $J_{\rho,x}$ related to links of the dual lattice take the
values $0,\:\pm 1,\: \pm 2$. We define current lines such that for
$J_{\rho,x}=0$ there is no line on the link, for $J_{\rho,x}=\pm 1$
there is one current line
in the positive or negative direction, respectively,
and for $J_{\rho,x}=\pm 2$
there are two lines in the positive or negative direction.
Because the $J_{\rho,x}$ are subject to (\ref{cc}) the same number
of lines must arrive at and depart from a site.

The current lines thus form connected sets which we call networks. The
topologically relevant ingredients of these networks are the vertices,
defined as the sites where at least two lines arrive (and depart), and the
edges, defined as the current lines connecting the vertices.

For networks of current lines it is intuitively clear whether
a network wraps around the torus in some direction or not. However, a
precise mathematical criterion remains to be given. It should be obvious
that cutting the network into loops is not allowed because (apart from being
highly nonunique) this would change the topology. In the following we point
out how the fundamental homotopy group $\pi_1$ can be used to obtain
the desired topological characterization.

The elements of $\pi_1$ correspond to equivalence classes of paths which
can be deformed continuously into each other and which all start and end at
the same point, called base point. A strategy to determine $\pi_1({\bf X},b)$
of a space ${\bf X}$ with a base point $b$ is to cover ${\bf X}$ by a suitably
dense network and to make use of the fact that the related edge path group is
isomorphic to $\pi_1$ \cite{span}. Analyzing the network then leads to
$\pi_1({\bf X},b)$. For the 4-dimensional torus considered here one gets
$\pi_1({\bf T}^4,b)={\bf Z}^4$ independently of the choice of $b$.

This motivates a related procedure which we propose for characterizing the
topology of the networks of current lines embedded in ${\bf T}^4$. For a
particular network ${\bf N}$ it exploits the observation that the analysis
provides the generators of $\pi_1({\bf T}^4,b)$ if ${\bf T}^4$ is suitably
covered by ${\bf N}$ , while it gives only those of a subgroup thereof if
${\bf N}$ does not wrap around in all directions. Thus one gets an appropriate
characterization by the (proper or improper) subgroup associated to ${\bf N}$.

To derive our rules we choose one vertex point of ${\bf N}$ to be the base
point $b$ and consider the set of all loops through $b$, i.e. of all
paths through ${\bf N}$ which start
and end at this point. We then use the fact that the group content of this
set is not changed if we perform mappings preserving the homotopy of all of
these paths. This in particular holds for a mapping by which one edge
shrinks to zero length. By a sequence of such mappings one finally can
shift all other vertices to $b$. One thus gets a bouquet of paths starting
and ending at the base point.

To perform this shrinking procedure in practice, we represent
any path on
${\bf T}^4$ by a vector which is the sum of the oriented steps along the
path. Thus a vector of this type is associated to each edge (and depends on
the starting point and the end point of the edge,
however, not on the particular
path it takes). Then a shrinking of one edge implies that the coordinates
of the moving vertex and the vectors of all other edges connected to this
vertex are to be modified appropriately.

For a network ${\bf N}$ with $K_0$ vertices and $K_1$ edges one obtains a
bouquet of $K=K_1 - K_0 +1$ loops on ${\bf T}^4$, which are related to elements
of $\pi_1({\bf T}^4,b)$ . The bouquet is described by a set of vectors of
the type introduced above, $\vec{s}_i$ with $i=1,\ldots,K$. If the $i$-th loops
winds around the torus $w_{ij}$ times in direction $j$ (including the sign),
the j-th component of the respective vector is $s_{ij}=w_{ij} L_j$ where $L_j$
denotes the lattice size and $j=0,1,2,3$. Thus one can equivalently use
the vectors $\vec{w_i}$ with components $w_{ij}$ to represent the bouquet.

The networks considered here have the additional properties of given path
orientation and of respecting current conservation at the vertices. This
reduces the allowed patterns. For the net current flow (\ref{nfn}) it
implies the relation $\vec{f}= \sum_i \vec{w}_i$ for the bouquet vectors,
which restricts the form of the bouquet matrix.

While the group content of the bouquet is unique, which particular loops
occur depends on the succession of the shrinking mappings chosen to form the
bouquet. Transformations between equivalent bouquets have to preserve homotopy
and to respect current conservation. We observe that elementary maps of this
type are ones in which three vectors, $\vec{w}_a$, $\vec{w}_b$, $\vec{w}_c$,
selected out of the bouquet, are replaced by $\vec{w}_a$,
$\vec{w}_b-\vec{w}_a$, $\vec{w}_c+\vec{w}_a$ (as one readily verifies
considering the partial network with two vertices from which, depending on
the edge selected for shrinking, the first or the second form arises).

Obviously these elementary maps correspond to steps of a modified Gauss
elimination procedure within the bouquet matrix $w_{ij}$, in which adding of
a row to another one requires to subtract it simultaneously from a further
row. Applying steps of this type $w_{ij}$ can be cast into a standard form
with rows $\vec{a}_1$, $\ldots$, $\vec{a}_r$, $\vec{t}$, $\vec{0}$, $\ldots$,
$\vec{0}$ where $\vec{a}_i\ne\vec{0}$ and where $r\le 4$ is minimal. Because
the entries of the matrix are integers and because divisions are not allowed
in the procedure, one in general remains with a triangular form of the
$\vec{a}_i$, for $r=4$ with $i$ elements which may differ from zero (while
in our application except for very few cases further reduction to $a_{ij}=
\pm \delta_{ij}$ occurs).

The pair form with rows $\vec{a}_1$, $-\vec{a}_1$, $\ldots$, $\vec{a}_r$,
$-\vec{a}_r$, $\vec{f}$, $\vec{0}$, $\ldots$, $\vec{0}$, which explicitely
exhibits $\vec{f}$, for $K\ge 2r+1$ is immediately obtained from the standard
form. For $\vec{f}=\vec{0}$ the number of nontrivial directions is $r$.
For $\vec{f}\ne\vec{0}$, rewriting the pair form for $K\ge 2r+3$ as
$\vec{a}_1$, $-\vec{a}_1$, $\ldots$, $\vec{a}_r$, $-\vec{a}_r$, $\vec{f}$,
$-\vec{f}$, $\vec{f}$, $\vec{0}$, $\ldots$, $\vec{0}$, it is seen that
the number of independent pairs out of $\vec{a}_1$, $-\vec{a}_1$, $\ldots$,
$\vec{a}_r$, $-\vec{a}_r$, $\vec{f}$, $-\vec{f}$ is the number of nontrivial
directions, which may be $r$ or, provided that $r+1\le 4$, also $r+1$.

\section{Monopole currents}
\setcounter{equation}{0}
\hspace{0.35cm}
There exist quite a number of contacts of current lines. We define their
number at a site by the number of lines arriving at the site
(or, equivalently, departing from it) minus one.
We find that their overall number per size is larger in the hot phase than
in the cold one, decreases with increasing $\lambda$, and shows little
dependence on $L$. The data in Table II give an overview of this.

The number of contacts in a network equals the number of links along
its lines minus the number of sites on its lines. We get a roughly linear
increase of the number of contacts with the size of a network (with some
increase of the fluctuations around the curve and of its slope with size).
For $\lambda=0$ this confirms an observation of Ref.~\cite{bls93}. For larger
$\lambda$ we find that the slope gets smaller. We see almost no $L$ dependence
of the slopes. Table II also contains mean numbers of contacts in a network per
network size (for networks larger than 19), which are seen to be similar to the
overall numbers.

In Figure 8a we depict the probability to find a network which is nontrivial
in at least one direction and in Figure 8b the probability to find one
nontrivial in four directions as functions of $\lambda$ along the transition
line for $L=8$. We have obtained the data for the hot (confining) and the
cold (Coulomb) phase by separating the $E$ histogram at the minimum between
the peaks (up to the $\lambda$ value where this has been still
possible). From Figure 8 it is obvious that the topological
characterization provides a signal for the phases.

We find that for $L=16$ this effect at $\lambda=0$ is already more
pronounced than it is for $L=8$ at $\lambda=-0.3$. The obvious reason for
this is that the peaks related to the phases become well separated, which
makes the signal for the phases rather perfect, the hot phase being
indicated by a network nontrivial in all directions, and the cold one by
the absence of nontrivial networks. This appears to be the generic situation
for larger lattices. It thus turns out that the topological characterization
provides an unambiguous signal for the phases.

It is useful to emphasize the difference between the topological
classification of the networks we have given here and the
``winding number'' $(1/L_{\mu})\sum_{x\in{\bf N}}M_{\mu,x}$, as defined in
Ref.~\cite{bls93}, which, because of current conservation, equals
the net current flow $f_{\mu}$ (\ref{nfn}).
Our topological characterization formalizes the intuitive notion
that a network of monopole loops wraps all around the torus,
i.e. that it contains oriented paths that allow one to go around
the torus and come back to the original point.  This can happen,
and thus give to the network a non-trivial topology,
even if the network carries no net current flow.  Indeed,
for $\lambda=0$ we also found numerically that the net current flow
is nonzero only in very rare cases for $L=8$ and not at all for
$L=16$. For larger $\lambda$ the fraction of such events increases.
Some of our data on the net current flow are reproduced in Table III.

Because of (\ref{nf}) $f_{\mu}\ne 0$ implies that more than one nontrivial
network occurs. From Table III it is seen that the case $\vec{f}\ne \vec{0}$
coincides indeed with the occurrence of
more than one nontrivial network. It also
shows that the number of these networks increases with $\lambda$.

In Figure 9 we present the probability to find networks being nontrivial
in 0 to 4 directions as function of $\lambda$ along the transition line
(without separating phases and thus allowing to cover the full range of
$\lambda$). For trivial networks within errors there is no dependence on
$\lambda$. The fraction of nontrivial networks being nontrivial in less
than four directions is seen to increases with $\lambda$. Thus there is
a $\lambda$ region where all of these structures become similarly important,
which is a further indication of the transition getting of second order.

Figure 10 shows the mean size of the largest network for $L=8$ as function
of $\lambda$ along the transition line (the statistical errors given are small
as compared to the fluctuations of sizes around the mean). The signal for
the phases is similar as in Figure 8, which reflects the fact that the
nontrivial network is large. Our data for $L=16$ show the same effect (for
$\lambda=0$ the mean size of the largest network in the hot phase is
12900(200) and in the cold phase it is 790(50)).
Figure 10 in addition reveals that
only the hot phase data change significantly with $\lambda$ (as can also
be observed for $\rho$ in Figure 7).

Figure 11 gives the average number $N(l)$ of trivial networks as a function
of their size $l$ for $L=16$ and $\lambda=0$ in the transition region.
The hot and cold phase data turn out to be rather similar. The distributions
within errors decrease with power laws, slightly faster for the hot phase.
The plots for the $L=8$ data look very similar, apart from the
numbers being smaller. They show very little depencence on $\lambda$. Table
IV, with the results of a fit of $N(l)$ versus $kl^{-z}$,
summarizes these findings.

The power law $N(l)\sim l^{-z}$ may be related to a fractal dimension
$D_{\mbox{\scriptsize{f}}}$. Assuming that the sum of lengths of networks
of size $l$ per volume, $lN(l)/V$, does not change under coarse graining,
by which it gets the form
$(l/b^{D_{\mbox{\scriptsize{f}}}})N(l/b^{D_{\mbox{\scriptsize{f}}}})/(b^DV)$,
one obtains $z= 1+D/{D_{\mbox{\scriptsize{f}}}}$. Inserting $z$ from Table
IV and $D=4$ it follows that $D_{\mbox{\scriptsize{f}}}$ is in the range
between 1.8 to 2.8 , i.e.~well below 4.

Our observations can be used to get insight into the mechanisms involved
in the phase transition.
We have noticed (cfr. Figures 7 and 10) that in the cold phase $\rho$ and the
size of the largest network show little dependence on $\lambda$, while in
the hot phase these quantities decrease strongly with $\lambda$. The
decrease of the latent heat (cfr. Figures 1, 3, and 5), due to the strong
correlation between $E$ and $\rho$ (apparent from Figures 6 and 7) essentially
only reflects the indicated behavior of $\rho$. We have also seen that for
trivial networks neither the distribution of their sizes (cfr. Table IV)
nor the probability to find them (cfr. Figure 9) show significant changes with
$\lambda$. Therefore, the quantity most affected by $\lambda$
must be the probability for the occurrence of nontrivial networks
in the hot phase.

Thus the following picture emerges. For
negative $\lambda$, there is typically one
large nontrivial network in the hot phase. With increasing $\lambda$
(and with the consequent
suppression of monopoles) there occurs a progressive thinning
of such network, which reduces its
size and the value of $\rho$.
Then increasingly it subdivides (Table III and Figure 9) and
breaks into smaller pieces.
The dynamics at fixed $\lambda$ may be illustrated in the following way. If
there is a very large nontrivial network, it will tend to thin out to reach
the size favored by the Boltzmann weight. On the torus it can, however, only
get gradually smaller to a minimal size beyond which it must break into
pieces. One possible explanation
of the first order nature of the transition for
small $\lambda$ would then be that, in absence of thinning, the
probability for the network to break up is low and
a substantial amount of energy is also required.
Therefore one gets the valley in the
two-peak distribution and a gap.

If instead of the torus ${\bf T}^4$ one considers the sphere ${\bf S}^4$,
because $\pi_1({\bf S}^n)$ for $n\ge 2$ only contains the neutral element,
the topological characterization no longer identifies distinct phases. To
illustrate the dynamics in that case one may again consider a very large
network. Now it can gradually get smaller without the above necessity to
break at some point. Thus there should be only one peak (located roughly
in the middle of the two-peak structure of a comparable torus).

The authors of Ref.~\cite{ln93}, which simulate
the system on the surface of a 5-dimensional cube, homeomorphic
to ${\bf S}^4$, observe indeed only one peak. Some caution appears
appropriate, however, because on smaller lattices the inhomogeneities of the
cube may cause smearing effects, that only the narrowing of the peak
for larger systems would exclude.

However appealing, the topological interpretation of the order of
the transition on the torus must face the notion
that first order transitions are bulk effects, in which boundaries play
no role. In view of our observation that for increasing lattice size the
topological characterization gets very clear, the disappearance of the
transition also on extremely large lattices is hard to imagine.
If one wishes to exclude the relevance of the boundary conditions
then the interpretation of our observations on the topological properties
of the networks would be that, although not crucial for the
order of the transition, they form an excellent diagnostic tool. They indicate
the occurrence of some type of percolation transition, whereby the monopole
loops condense into a network pervading all
of (4-dimensional) space.  Also, the discrepancy
of the results of ~\cite{ln93} with such picture would remain to be explained.
If the topological properties of the networks are, instead,  intimately
connected to the nature of the transition, this rises
the question if such transitions, which certainly are of interest in models,
could have physical implications, too.

\section{Conclusions}
\hspace{0.35cm}
Adding a monopole term to the action has allowed us to set up a powerful
simulation algorithm, to study the extended theory, and to extract
the underlying
mechanisms of the phase transition. We have found that the strength of the
first order transition decreases with the weight of the added term
in such a way that
the transition ultimately gets of second order. We have presented
detailed data on the properties of the system
in this context. In order to be able to analyze the occurring configurations
appropriately, we have worked out the topological characterization of networks
of current lines. From our analysis we have obtained detailed results on
these networks. In particular, we have found that their topological properties
signal the phases.

\section*{Acknowledgements}
One of us (W.K.) wishes to thank the Physics Department of Boston University
for kind hospitality during his visits.
He also thanks Thomas Kerler for helpful
discussions on topology. This work has been supported in part by the Deutsche
Forschungsgemeinschaft through Grants Nos. Ke 250/7-1, 7-2, 9-1, and
11-1 and by the United States Department of Energy under Grant
No. DE-FG02-91ER40676.
The computations have been done on the CM5 of the Center for Computational
Science of Boston University and since recently also on the CM5 of the GMD
at St.~Augustin.

\newpage
\begin{center}

{\bf Table I}\\
$\beta_{\mbox{\scriptsize{C}}}$ of phase transition for $L=8$ ;\\
$\beta$ and $g(\lambda)$ of simulations.\\
\vspace{1mm}
\begin{tabular}{|c|c|c|c|}
\hline
$\lambda$ & $\beta_{\mbox{\scriptsize{C}}}$ & $\beta$ & $g(\lambda)-g(-0.3)$ \\
\hline
  -0.30   &   1.1786(1) &  1.1785  &  $0.00000\times 10^{0}$ \\
  -0.25   &   1.1501(1) &  1.1501  &  $1.74914\times 10^{2}$ \\
  -0.20   &   1.1217(1) &  1.1217  &  $3.54865\times 10^{2}$ \\
  -0.15   &   1.0932(1) &  1.0932  &  $5.40888\times 10^{2}$ \\
  -0.10   &   1.0647(1) &  1.0646  &  $7.33197\times 10^{2}$ \\
  -0.05   &   1.0361(1) &  1.0361  &  $9.30199\times 10^{2}$ \\
   0.00   &   1.0075(1) &  1.0074  &  $1.13491\times 10^{3}$ \\
   0.05   &   0.9787(1) &  0.9788  &  $1.34483\times 10^{3}$ \\
   0.10   &   0.9496(1) &  0.9498  &  $1.56490\times 10^{3}$ \\
   0.15   &   0.9203(1) &  0.9205  &  $1.79466\times 10^{3}$ \\
   0.20   &   0.8908(1) &  0.8909  &  $2.03454\times 10^{3}$ \\
   0.25   &   0.8609(1) &  0.8610  &  $2.28503\times 10^{3}$ \\
   0.30   &   0.8304(1) &  0.8306  &  $2.54875\times 10^{3}$ \\
   0.35   &   0.7995(1) &  0.7998  &  $2.82532\times 10^{3}$ \\
   0.40   &   0.7680(1) &  0.7685  &  $3.11653\times 10^{3}$ \\
   0.45   &   0.7359(1) &  0.7364  &  $3.42644\times 10^{3}$ \\
   0.50   &   0.7028(1) &  0.7034  &  $3.75730\times 10^{3}$ \\
   0.525  &   0.6860(1) &  0.6864  &  $3.93287\times 10^{3}$ \\
   0.55   &   0.6688(2) &  0.6693  &  $4.11274\times 10^{3}$ \\
   0.575  &   0.6512(2) &  0.6515  &  $4.30418\times 10^{3}$ \\
   0.60   &   0.6335(2) &  0.6337  &  $4.49916\times 10^{3}$ \\
\hline
\end{tabular}

\newpage

{\bf Table II}\\
Contacts per size in units of 100.\\
\vspace{1mm}
\begin{tabular}{|c|c|c|c|c|}
\hline
   $L$    & $\lambda$  &  phase    &  overall  &  network \\
\hline
    8     &   -0.3     &  cold     &  6.2(1)   &  9.7(2) \\
          &            &  hot      & 11.3(1)   & 11.7(2) \\
    8     &    0.0     &  cold     &  5.5(1)   &  8.4(1) \\
          &            &  hot      &  7.9(1)   &  8.7(1) \\
    8     &    0.3     &  cold     &  4.4(1)   &  6.7(1) \\
          &            &  hot      &  5.9(1)   &  7.0(1) \\
    8     &    0.6     &           &  3.9(1)   &  5.6(1) \\
    8     &    0.9     &           &  3.0(1)   &  4.5(1) \\
   16     &    0.0     &  cold     &  5.1(1)   &  8.2(1) \\
          &            &  hot      &  7.1(1)   &  8.2(1) \\
   16     &    0.6     &           &  3.4(1)   &  5.5(1) \\
\hline
\end{tabular}

\vspace{1cm}

{\bf Table III}\\
Probabilty for $\vec{f}\ne \vec{0}$ and for more than\\
one nontrivial network in units of 100.\\
\vspace{1mm}
\begin{tabular}{|c|c|c|c|c|c|}
\hline
   $L$ & $\lambda$ & $\vec{f}\ne \vec{0}$ & \# = 2 & \# = 3 & \# = 4 \\
\hline
    8  &  -0.3  &  0.6(4)   &  0.6(4)   &  0.0(2) &  0.0(2) \\
       &   0.0  &  1.6(5)   &  1.6(5)   &  0.0(2) &  0.0(2) \\
       &   0.3  &  6.6(1.3) &  6.6(1.3) &  0.0(2) &  0.0(2) \\
       &   0.6  & 14.6(1.5) & 12.7(1.5) &  1.4(4) &  0.5(3) \\
       &   0.9  & 19.3(3.3) & 15.9(2.7) &  3.0(5) &  0.4(3) \\
   16  &   0.0  &  0.0(3)   &  0.0(3)   &  0.0(3) &  0.0(3) \\
       &   0.6  & 12.0(2.7) & 12.0(2.7) &  0.0(3) &  0.0(3) \\
\hline
\end{tabular}

\newpage

{\bf Table IV}\\
$z$ and $k$ from fits $kl^{-z}$ to the probability\\
for trivial networks of size $l$.\\
\vspace{1mm}
\begin{tabular}{|c|c|c|c|c|}
\hline
   $L$    & $\lambda$  &  phase    &   z       &      k      \\
\hline
    8     &   -0.3     &  cold     &  2.46(2)  &   1164(42)  \\
          &            &  hot      &  3.22(3)  &   2059(110) \\
    8     &    0.0     &  cold     &  2.48(2)  &   1168(27)  \\
          &            &  hot      &  3.01(2)  &   1814(60)  \\
    8     &    0.3     &  cold     &  2.48(2)  &    995(3)   \\
          &            &  hot      &  2.87(2)  &   1494(43)  \\
    8     &    0.6     &           &  2.66(1)  &   1158(21)  \\
    8     &    0.9     &           &  2.65(2)  &   1007(36)  \\
   16     &    0.0     &  cold     &  2.41(1)  &  15430(130) \\
          &            &  hot      &  2.85(1)  &  25830(180) \\
   16     &    0.6     &           &  2.63(1)  &  19710(250) \\
\hline
\end{tabular}

\end{center}
\newpage

\section*{Figure captions}
\begin{tabular}{rl}
Fig. 1. & Distribution $P(E)$ in the transition region\\
        & on lattices with $L=8$ (rhombs) and $L=16$ (crosses),\\
        & (a) for $\lambda=0$ and (b) for $\lambda=0.6$~.\\
Fig. 2. & Distributions $P(E)$ for $\lambda=0.6$ and $L=16$,\\
        & for $\beta=0.6432$ (rhombs), $\beta=0.6428$ (crosses),
          $\beta=0.6424$ (squares).\\
Fig. 3. & Distribution $P(E,\lambda)$ for $L=8$\\
        & with $\beta$ in the transition region.\\
Fig. 4. & Location of phase transition in ($\beta,\lambda)$ space\\
        & for $L=8$, $\beta$ versus $\lambda$.\\
Fig. 5. & Latent heat for $L=8$ as function of $\lambda$\\
        & at the transition line.\\
Fig. 6. & Distribution $P(E,\rho)$ for $L=8$ and $\lambda=-0.3$\\
        & in the transition region.\\
Fig. 7. & $P(E,\rho)$ for $L=8$ in the transition region,\\
        & for $\lambda=-0.3,\:0,\:0.3,\:0.6,\:0.9$ (giving the\\
        & distributions from left to right, respectively)\\
Fig. 8.  & Probability for a network in cold (rhombs) and hot (crosses) phase\\
         & as function of $\lambda$ for $L=8$,\\
         & (a) being nontrivial in at least one direction,\\
         & (b) being nontrivial in four directions.\\
Fig. 9.  & Probability for a network being nontrivial in 0 to 4 directions\\
         & as function of $\lambda$ for $L=8$.\\
Fig. 10. & Mean size of largest network in cold (rhombs) and hot (crosses)
           phase\\
         & as function of $\lambda$ for $L=8$.\\
Fig. 11. & Number $N(l)$ of trivial networks as function of size $l$\\
         & for $L=16$ and $\lambda=0$,\\
         & (a) cold phase, (b) hot phase.\\
\end{tabular}

\end{document}